\documentclass[preprint,prl]{revtex4}
\usepackage{epsfig}
\usepackage{floatflt}
\usepackage{graphicx}
\usepackage{amsmath}
\usepackage{setspace}
\begin{document}

\title{Evidence of growing spatial correlations at the glass transition from nonlinear response experiments}

\author{C. Crauste-Thibierge$^1$}
\author{C. Brun$^1$}
\author{F. Ladieu$^{1 \star}$}
\author{D. L'H\^ote$^{1 \star}$}
\author{G. Biroli$^2$}
\author{J-P. Bouchaud$^3$}
\email{francois.ladieu@cea.fr, denis.lhote@cea.fr}

\affiliation{$^1$ SPEC (CNRS URA 2464), DSM/IRAMIS CEA Saclay, Bat.772, F-91191 Gif-sur-Yvette  France}
\affiliation{$^2$ Institut de Physique Th\'eorique, CEA, (CNRS URA 2306), 91191 Gif-sur-Yvette, France}
\affiliation{$^3$ Science $\&$ Finance, Capital Fund Management, 6, Bd. Haussmann, 75009 Paris, France}

\date{\today}

\begin{abstract}
The ac nonlinear dielectric response $\chi_3(\omega,T)$ of glycerol was measured close to its 
glass transition temperature $T_g$ to investigate the prediction that supercooled liquids respond in an increasingly non-linear way as the dynamics slows down (as spin-glasses do). We find that $\chi_3(\omega,T)$ indeed displays several non trivial features. It is peaked as a function of the frequency $\omega$ and obeys scaling as a function of $\omega \tau(T)$, with $\tau(T)$ the relaxation time of the liquid. The height of the peak, proportional to the number of dynamically correlated molecules 
$N_{corr}(T)$, increases as the system becomes glassy, and $\chi_3$ decays as a power-law of $\omega$ over several decades beyond the peak. These findings confirm the collective nature of the glassy dynamics and provide the first direct estimate of the $T$ dependence of $N_{corr}$.

\vskip 5.0cm
\large{{\it Accepted for publication in Physical Review Letters}}

\end{abstract}

\maketitle

Most liquids undergo a transition to an amorphous solid state, the glass, when the temperature $T$ decreases to their glass transition temperature $T_g$. This transition is similar for a vast variety of systems such as polymeric, colloidal, molecular liquids. This ubiquity echoes the universality of critical phenomena, where the emergence of long range order makes irrelevant most of the microscopic details, yielding the same critical behavior in various systems. This is why it has been argued for a long time that some collective effects should be associated to the glass transition, ultimately related to the proximity of a phase transition~\cite{Dyre}. An immediate problem with this idea is that standard equilibrium correlation and response functions remain those of a normal liquid when $T \to T_g$ \cite{Debenedetti}. In the two past decades, however, the heterogeneous nature of the dynamics close to $T_g$ was established \cite{HoleBurning,RevRichert}, and a length scale associated to these dynamic heterogeneities was estimated \cite{RMN}. The increase of the number of correlated molecules $N_{corr}$ when $T$ decreases towards $T_g$ is expected to explain the main aspect of the glass transition, i.e. the huge increase of the relaxation time; $N_{corr}(T)$ has therefore become a central concern in the field. 
The first estimate of this dependence \cite{science, JCP1, PRE} relied on the temperature derivative of a two-point correlation function. Its relation with the number of dynamically correlated molecules $N_{corr}(T)$ is however
ambiguous because it involves a temperature dependent prefactor difficult to estimate and control \cite{JCP1,NoteJCP1}.
Furthermore, this estimator leads to a low-temperature divergence of $N_{corr}(T)$ for a purely Arrhenius system, for which no collective behaviour is expected, at least at first sight \cite{JCP1}. An indisputable experimental estimate of the $T$-dependence of the dynamical correlation volume is thus still lacking. In this work, we fill this gap by measuring the anomalous increase of a physical susceptibility.  It leads to a {\it direct} estimate of $N_{corr}(T)$, up to a numerical prefactor that is now temperature independent. Our findings show that supercooled liquids respond in an increasingly non-linear way approaching the glass transition. For many systems, {\it e.g.} spin glasses, such an increase is related to criticality. In the present case, it suggests that an underlying phase transition could possibly be present as well, although it does not necessarily imply it.

The spin glass transition taught us that two-point functions can be blind
to the amorphous long range order and that this critical behavior is revealed in particular by the third order nonlinear magnetic susceptibility \cite{BY,LLevy}. 
It was recently argued~\cite{BB-PRB,Tarzia} that the frequency and temperature dependent nonlinear dielectric susceptibility 
$\chi_3(\omega,T)$ in supercooled liquids plays a role similar to the nonlinear 
magnetic susceptibility in spin glasses and that its increase when $T$ decreases
would be a signature of the incipient long range amorphous order. 
Accordingly, $\chi_3(\omega,T)$ should display a peak for $\omega \tau(T) \sim 1$ whose height 
grows as one approaches $T_g$. Here $\tau(T)$ is the relaxation time of the supercooled liquid
which increases rapidly as $T$ decreases toward $T_g$. Contrary to spin glasses, however, 
$\chi_3(\omega = 0,T)$ should remain trivial since at long times glasses still behave as disordered liquids \cite{Tarzia}. One expects that when $\tau (T)$ is 
large, $\chi_3$ takes the following scaling form:
\begin{equation}\label{scaling}
\chi_3 (\omega,T) \approx \frac{\epsilon_0 (\Delta \chi_1)^2 a^3}{k_BT} N_{corr}(T) \, {\cal H}\left(\omega \tau\right),\qquad
\end{equation}
where $N_{corr}(T)$ is the $T$-dependent average number of dynamically correlated molecules, 
$\Delta \chi_1$ = $\chi_1(\omega = 0) - \chi_1(\omega \rightarrow \infty)$ is the part of the static linear susceptibility corresponding to the slow relaxation process we consider,
$a^3$ the volume occupied by one molecule, and 
${\cal H}$ a certain complex scaling function that goes to zero {\it both} for small and large arguments. The humped shape of $|{\cal H}\left(\omega \tau\right)|$ is a {\it distinctive feature} of the glassy correlations. Indeed, in the `no correlation case' \cite{Supplement, perpignan}, $\chi_3(\omega, T)$ is given by the prefactor of Eq. (\ref{scaling}) times a function that, for all $T$, reaches its maximum value at $\omega =0$. Furthermore $N_{corr}(T)$ does not show any temperature dependence in that case.

We have devised an experiment to measure, for the first time, $\chi_3(\omega, T)$ of a classic glass-former (glycerol), via third harmonics measurements. Our findings are in good agreement with the predictions of Eq. (\ref{scaling}): the scaling, the growth of the number of dynamically correlated molecules with decreasing $T$, the humped shape of $|{\cal H}|$ and its power-law dependence for $\omega \tau \gg 1$ are all confirmed experimentally. The anomalous increase of $\chi_3$ is compatible with the existence of incipient critical fluctuations when $T \to T_g$, even if our measurements are too far from the possible critical point to be a clue of criticality.

Our measurements were performed at temperatures between $T_g + 4$ K and $T_g + 35$ K, with $T_g \approx 190$ K. The linear susceptibility $\chi_1 (\omega) = \chi_1'(\omega)+ i \chi_1''(\omega)$ quantifies the first harmonic response of polar molecules to a periodic excitation field $E e^{-i\omega t}$. The polarization, $P(\omega)e^{-i\omega t}$, of the system is given to first order by $P(\omega) = \epsilon_0 \chi_1(\omega) E$. We define the relaxation time $\tau(T)$ as the inverse of the frequency $f_{\alpha}$ where $\chi_1''$ is maximum. The nonlinear response is dominated by $\chi_3(\omega)$ which gives to first order the magnitude of the third harmonics response to the field at pulsation $\omega$ \cite{Thibierge, Supplement}, $P(3\omega)e^{-3i\omega t} = \epsilon_0\chi_3(\omega)[Ee^{-i\omega t}]^3$ (the second harmonics is zero because of the $E \to -E$ symmetry). While $\chi_1(\omega,T)$ has been widely studied in supercooled liquids \cite{Lunkenheimer,Rossler,Autres1w}, the $T$ and $\omega$ dependence of $\chi_3$ has not been measured so far, for any liquid, glassy or not \cite{Wu}.   

In a closed cell, two high purity glycerol samples were prepared between gold plated copper electrodes 2 cm in diameter, with 
mylar$^{\textregistered}$ spacers ensuring inter-electrode distances of $19$ $\mu$m and $41$ $\mu$m \cite{Supplement, Thibierge}. The cell was placed in a cryogenerator, and the temperature $T$ of the samples was regulated with a precision of $50$ mK. A low harmonic distortion voltage source yielded a field $E \sim 7\times 10^5$ V/m (rms) in the thinnest capacitor, and $\chi_3(\omega , T)$ was obtained from the measured third harmonics current $I(3 \omega)$. This nonlinear signal was $10^{-7}$ to $10^{-5}$  of the linear one, well below the harmonic distortion of electronic devices. To get rid of them, we used a high sensitivity method based on a bridge containing the two glycerol capacitors~\cite{Thibierge} . 

\begin{figure} 
\includegraphics[width=16.2cm,height=11.7cm]{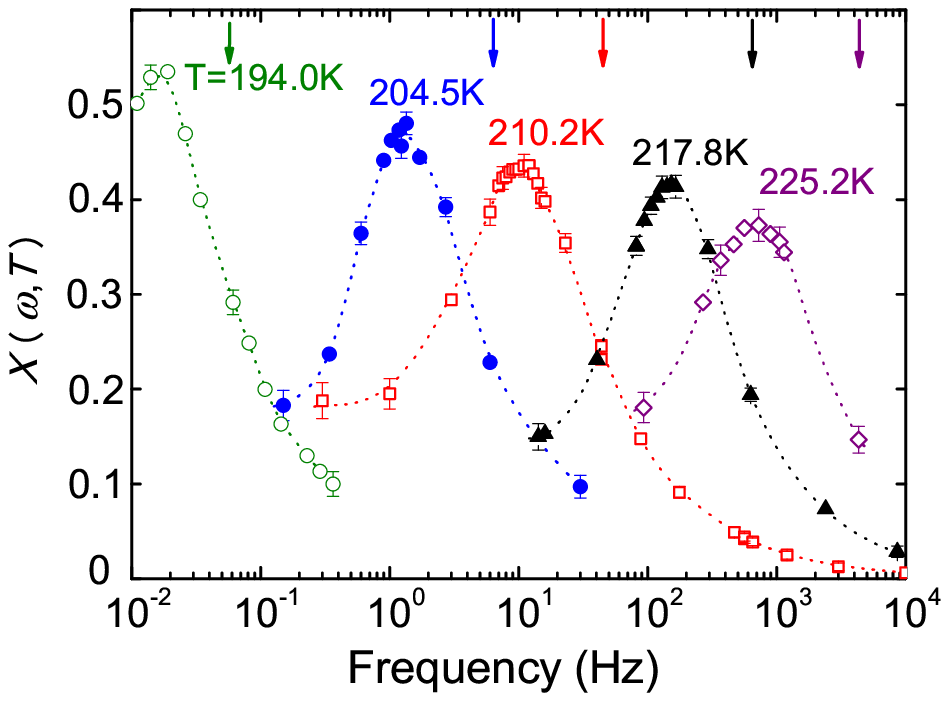} 
\caption{\label{fig1} (Color online) For each of the five temperatures labelling the curves, $X(\omega,T)$ = $\left|\chi_3(\omega,T)\right| \times k_B T/[\epsilon_0 (\Delta \chi_1)^2 a^3]$ is given as a function of frequency $f$ = $\omega /2\pi$. The arrows indicate the five relaxation frequencies $f_{\alpha}(T)$, for which $\chi_1''(\omega)$ is maximum. The dashed lines are guides to the eyes.} 
\end{figure}

Figure 1 shows, for five temperatures, the $\omega$ dependence of $\mathnormal{X}(\omega,T)=\left|\chi_3(\omega,T)\right|\times k_B T/[(\Delta \chi_1)^2 a^3 \epsilon_0]$. According to Eq. (\ref{scaling}), this quantity should be $N_{corr}(T) \left|{\cal H}(\omega \tau)\right|$. Two striking results, confirming the predictions of Eq. (\ref{scaling}), can be seen in Fig. 1. 
First, $\mathnormal{X}(\omega,T)$ is actually peaked at a frequency $f^*$ of the order of $1/\tau(T)$; more precisely $f^* \simeq 0.21/\tau(T)$. Second, the height of the peak, which should be proportional to $N_{corr}(T)$, increases significantly as $T$ decreases. The small value of $\mathnormal{X}(\omega,T)$ can be understood from calculations of $\chi_3(\omega,T)$ for molecules undergoing independent rotational Brownian motion \cite{Supplement, perpignan}, which should be accurate for simple supercooled liquids at high $T$. They indicate that for the `no correlation' case, $\mathnormal{X}(\omega,T) \le 0.2$.

\begin{figure} 
\includegraphics[width=16.2cm,height=11.7cm]{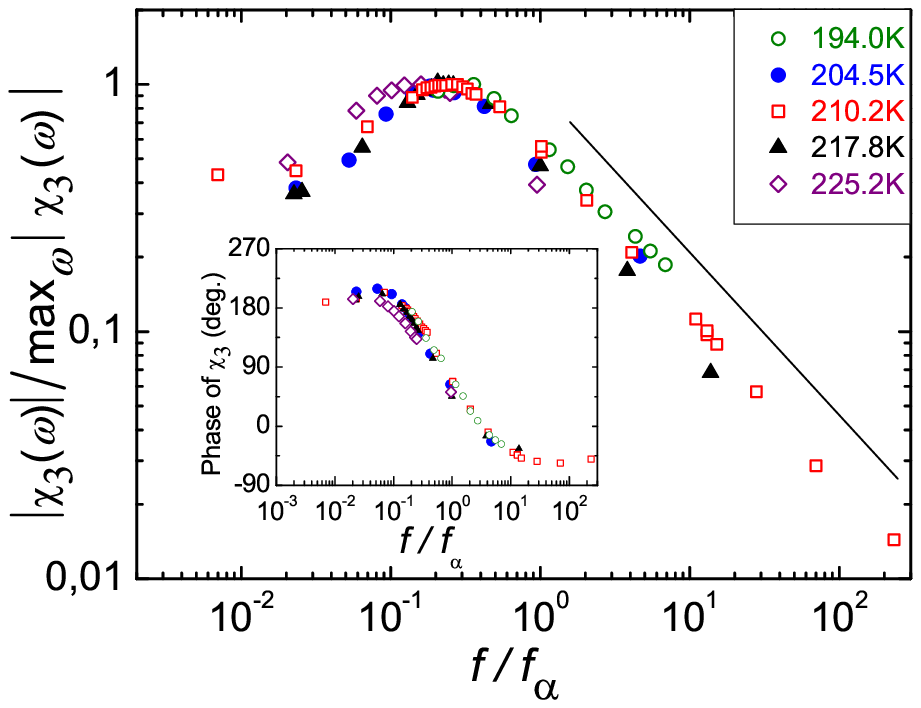} 
\caption{\label{fig2} (Color online) The quantity $\left| \chi_3(\omega,T)\right|$ at five temperatures (same data and symbols as in Fig. 1),  normalized to its maximum value at each $T$, plotted as a function of the ratio $f/f_{\alpha}(T)$. The straight line is a power-law with an exponent -0.65. Inset: corresponding phase of $\chi_3(\omega,T)$.} 
\end{figure} 

To what extent is the scaling predicted by Eq. (\ref{scaling}) verified by $\chi_3(\omega ,T)$? Fig. 2 shows the same data as in Fig. 1 but plotted as a function of $\log(f/f_{\alpha})$  with $f_{\alpha}=\tau(T)^{-1}$, and normalized by their maximum at each $T$. We also show the phase of $\chi_3$. Both the modulii and the phases collapse fairly well on a single master curve, as predicted by Eq. (\ref{scaling}). A weak departure from scaling occurs at low $\omega$ for the highest $T$. This is not surprising: scaling
should not hold far above $T_g$, where the dynamical correlations become short-ranged \cite{BB-PRB,Tarzia}. Furthermore, the nonlinear response results both from trivial dielectric saturation effects \cite{Richert07a, Bottcher,perpignan} (always present  for $\omega \tau < 1$), and from the non trivial dynamical correlations contribution. As the latter vanishes when $\omega$ $\rightarrow$ 0 (\cite{BB-PRB, Tarzia}), the contribution of the trivial saturation effect should dominate at high $T$ and low $\omega$. The observed departure from scaling can thus be explained by the different $T$ and $\omega$ dependencies of the two contributions. 

\begin{figure} 
\includegraphics[width=16.2cm,height=11.7cm]{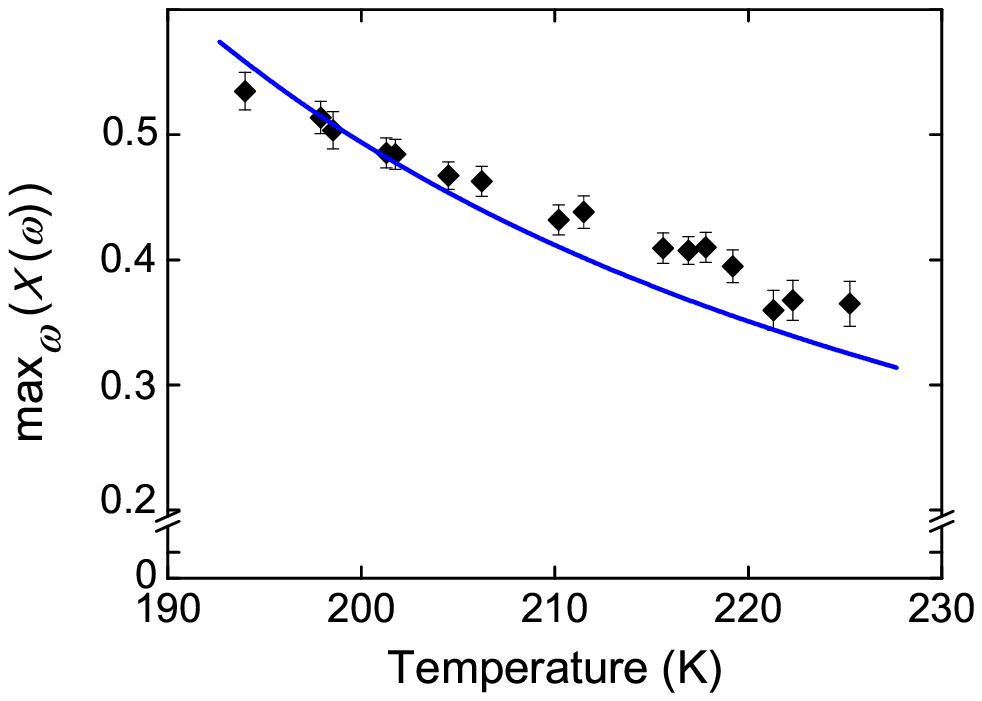} 
\caption{\label{fig3} (Color online) The measured quantity $\textrm{max}_{\,\omega}\left[X(\omega,T)\right]$ = $\textrm{max}_{\,\omega}(\left|\chi_3(\omega,T)\right|) \times k_B T/[\epsilon_0 (\Delta \chi_1)^2 a^3]$ plotted vs. temperature. The continuous line is the number of correlated molecules estimated from $T \chi_T$ (see \cite{PRE}) with an arbitrary normalization chosen so that it coincides with the experimental points at $202$ K.} 
\end{figure}

Figure 3 gives the $T$ dependence of the maximum value of $\mathnormal{X}(\omega,T)$, reached for $\omega$ = $\omega^*$ = $2\pi f^*$. Since 
scaling is well obeyed, Eq. (\ref{scaling}) tells us that this maximum value is proportional to $N_{corr}(T)$. A clear increase of this quantity as $T$ decreases toward $T_g$ is visible in Fig. 3. This is one of the main results of our study, and a strong experimental evidence of a growth of the dynamical correlation length close to $T_g$. We now investigate how our results are related to another estimate of the size of the dynamical heterogeneities, using the temperature derivative of $\chi_1(\omega)$.

In \cite{JCP1, PRE}, the $T$-derivative of the two-body correlation $C(t)$ (or of $\chi_1(\omega)$) was introduced as a new response function $\chi_T(t)=\partial C(t)/\partial T$. It turns out that $T\chi_T(t)$ is also proportional to a dynamical correlation volume, but with an unknown, and possibly $T$ dependent prefactor~\cite{JCP1,NoteJCP1}. The line in Fig. 3 is the number of correlated molecules estimated from $\max_{\,\omega} (T \partial (\chi_1'(\omega)/ \Delta \chi_1 ) /\partial T)$ where, due to the unknown value of ${\cal H}$, the normalization was chosen such that it coincides with $\mathnormal{X}(\omega^*,T=202 K)$. We see that the $T$ dependencies of these two quantities are close. However, whereas the increase of $\mathnormal{X}(\omega^*,T)$ is a proof of a growing correlation length, one needs some extra assumptions to infer this from $\chi_T$ \cite{JCP1}. That $\chi_3$ and $\chi_T$ lead to a similar $T$ dependence for $N_{corr}(T)$ suggests that these assumptions are indeed warranted (at least for glycerol) and validates the simpler procedure advocated in~\cite{science}, and used extensively in \cite{PRE}, to extract $N_{corr}(T)$. 

\begin{figure} 
\includegraphics[width=16.2cm,height=11.7cm]{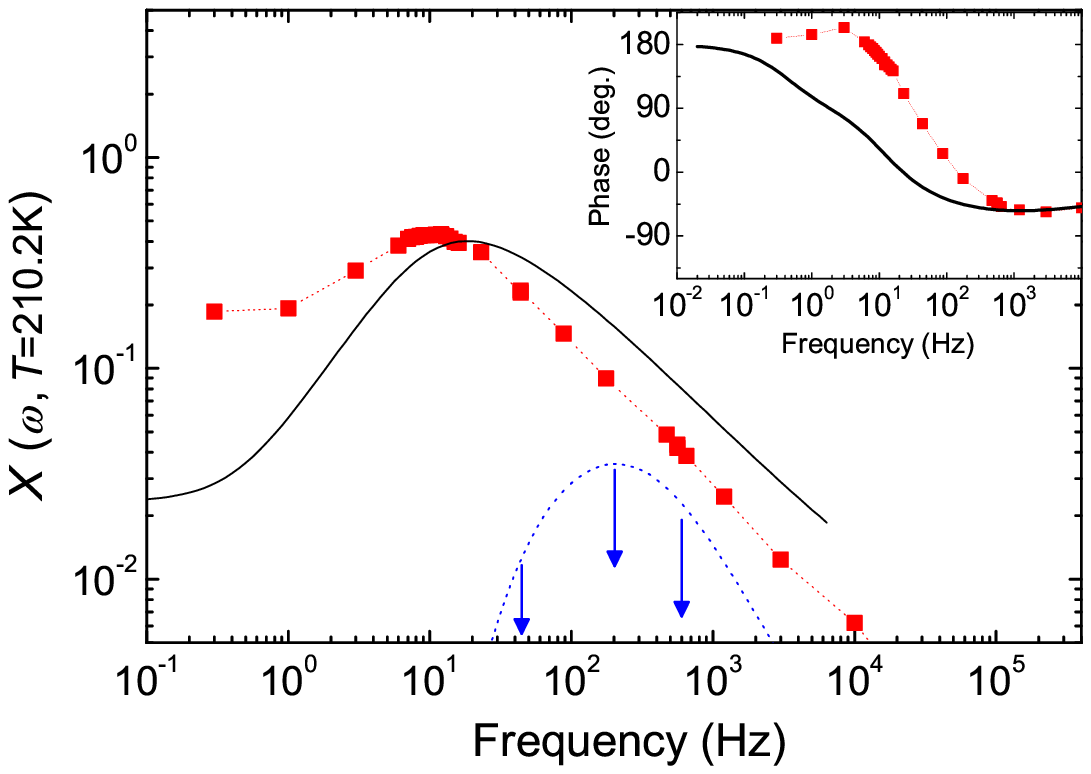} 
\caption{\label{fig4} (Color online) The measured quantity $X(\omega,T)$ at 210.2 K (filled squares with dashed line) is compared to the same quantity calculated by using Eq. (\ref{scaling2}) with $\kappa \approx 7.6 \times 10^{-17}$ K(m/V)$^2$). Dotted line: Calculated heating contribution (upper limit, the arrows indicate the possible reduction \cite{Supplement}). Inset: The phase of $\chi_3(\omega,T)$ at 210.2 K (same symbols as for the main figure, dotted line not shown).} 
\end{figure} 
  
A precise identity relating $\chi_T$ and $\chi_3$ has been obtained in~\cite{Tarzia}. It holds at low enough $\omega$ when the linear response 
depends only on external parameters ($T$, density, electric field, \ldots) via the dependence of $\tau (T)$ on these parameters, a property called time temperature superposition (TTS), and
reads: 
\begin{equation}\label{scaling2}
\chi_3 (\omega) \approx \kappa \hat \chi_T (2\omega)\qquad \omega \tau \ll 1  \ ,
\end{equation}
where $\kappa$ is independent of $\omega$ and $\hat \chi_T$ is the Fourier transform of $\chi_T$. Corrections to this relation in the regime $\omega \tau \ll 1$ are due to weak violations of TTS and to the additional term that corresponds to $\omega$ independent non-cooperative degrees of freedom. The continuous line on Fig. 4 presents $ \mathnormal{X}(\omega,T) $ calculated from the linear susceptibility, according to Eq.  (\ref{scaling2}) with $\kappa = 7.6 \times 10^{-17}$ K(m/V)$^2$. This value of $\kappa$ was chosen to be consistent with Fig. 3, i.e. by demanding that the maximum over $\omega$ of $\kappa \left|\hat \chi_T (2\omega)\right|$ corresponds to $X=0.40$ which is the value of the continuous line of Fig. 3 for $T=210.2K$. Although Eq. (\ref{scaling2}) should only be valid at small $\omega$, we see that it reproduces qualitatively the overall behavior of the data at $210.2$ K. At low frequency, $\omega \tau < 0.2$, the discrepancy can be understood by the contribution of the $\omega$ independent trivial saturation effect noted
above. For $\omega \tau > 1$, our data exhibit a clear power-law behavior with a fitted exponent $-0.65 \pm 0.04$, see also Fig. 2, which is close to that 
describing the decay of $\hat \chi_T(2\omega)$. This behavior is predicted by Mode-Coupling Theory~\cite{Tarzia}, and our experimental exponent is compatible with the MCT prediction $-b \simeq -0.6 $ for glycerol \cite{Lunkenheimer2}. Note however that MCT should in fact be relevant only at temperatures much higher than the ones we focused on. Hence, the existence of a power-law regime for large $\omega$ with identical exponents for both $\chi_1$ and $\chi_3$ appears to be a generic property, valid outside the MCT regime. We think that this might be related to the possible incipient criticality of the system.

We now address the question of a possible heating contribution to our results. 
Such effects would come from the fact that applying the field $E(t)=E \cos(\omega t)$ across the glycerol sample leads to a dissipated power per unit volume $p(t) \propto \omega \chi''_{1}(\omega) E^2 (1+ \cos(2 \omega t-\phi))$ (see \cite{Supplement}). 
The resulting temperature increase is the sum of a constant term and a term $\delta T_{2 \omega}(t)$, oscillating at $2 \omega$. This  $\delta T_{2 \omega}(t)$ oscillation leads to a modulation of the linear susceptibility $\delta \chi_1(\omega)(t)= (\partial \chi_1(\omega)/\partial T)\delta T_{2 \omega}(t)$, thus to a $3\omega$ modulation of the polarization $P(t) \propto \chi_1 E $. We calculated precisely this $3 \omega$ heating contribution using the heat propagation equation~\cite{Supplement}. This calculation gives an {\it upper limit} to the heating effects contribution, in particular for $f \ge f_{\alpha}$, because the finite relaxation time $\tau$ of the molecules prevents them from following instantaneously the temperature oscillation $\delta T_{2\omega}(t)$ (as already advocated by Richert \textit{et al.} \cite{Richert08}). A representative example of these calculations is given on Fig. 4 : the heating effects are negligible in most of the frequency range of interest, in particular around the $\chi_3(\omega)$ peak. Thus, our estimate of $N_{corr}(T)$ and the power-law dependence of $\chi_3(\omega)$ are not affected by heating.

Our results should allow significant progress of models which predict or assume dynamical correlations in supercooled liquids. Such models should not only reproduce the
increase of the correlated volume we found as $T$ $\rightarrow$ $T_g$ (see Fig. 3), but they should also account for the magnitude and shape of $\chi_3(\omega,T)$, given by the ${\cal H}$ function in Eq. (\ref{scaling}), which carries an important information on the physics of the dynamical correlations~\cite{BB-PRB,Tarzia}. 
Recently, Richert {\it et al.} put forward a phenomenological model~\cite{Richert07a,Richert08,Richert06} which accounts for their nonlinear susceptibility data at $1\omega$. In this model, the supercooled liquid is thought of as a collection of ``Debye-like'' dynamical heterogeneities ({\sc dh}), each of them having its own relaxation time $\tau_{dh}$. The model posits that the electrical power absorbed by each {\sc dh} raises its temperature above that of the phonon bath. A non-trivial and crucial assumption is that heat exchange between the slow degrees of freedom and the phonon bath is set by the local relaxation time $\tau_{dh}$, and not by a microscopic vibration time. As a result, each {\sc dh} has its own fictive temperature with a dc and an ac component. Again, the ac component $\delta T_{dh,2\omega}(t)$ leads to a $3\omega$ ``heating'' contribution to the polarization $P(t)$. Such a contribution to $\chi_3$ should however not be considered to be in competition with the critical dynamical correlations related to the glassy dynamics~\cite{BB-PRB,Tarzia}. As this model assumes {\it a priori} the existence of dynamical heterogeneities, it should be seen as a phenomenological description of the influence of dynamical correlations on the nonlinear susceptibility. We calculated the prediction of this model for the $3\omega$ nonlinear response and found that $\chi_3(\omega)$ is indeed peaked at a frequency of the order of $1/\tau(T)$. However, some assumptions are needed to generalize the model to the $3\omega$ response. These assumptions influence the magnitude and position of the peak and deserve further scrutiny. A detailed comparison of this model to our data will be presented in a subsequent paper. In any case, we believe that for a proper prediction of $\chi_3(\omega,T)$ close to $T_g$, a theory of supercooled liquids able to account for dynamical correlations is required.
 
To conclude, we have provided the first direct experimental evidence that a supercooled liquid responds in an increasingly non-linear way approaching the glass transition. By measuring the frequency dependent third harmonics response $\chi_3(\omega, T)$ to a periodic electric field, which is tightly related to the dynamical correlation length, we showed that the number of correlated molecules increases as $T$ decreases towards $T_g$, confirmed the validity of previous estimates, and found that $\chi_3$ scales as a function of $\omega \tau$.
This opens a new path for probing the spatial correlations in both fragile and strong supercooled liquids and in the aging regime of glasses and spin-glasses, by systematic studies of non-linear responses. Future investigations along these lines might help to unveil the possible critical nature of the glass transition.

{\bf Acknowledgements} 

We thank R. Tourbot for realizing the experimental cell, P. Pari for cryogenics, S. Nakamae for her careful reading of the manuscript. We acknowledge interesting discussions with C. Alba-Simionesco, A. Lef\`evre, R. Richert, M. Tarzia, and support by ANR grant DynHet.

\pagebreak

\renewcommand{\figurename}{Fig. S}
\setcounter{figure}{0}
\par
{\bf Electronic Physics Auxiliary Publication Service: Supplementary Information for : }

\begin{center}
	\textbf{\large{Evidence of growing spatial correlations at the glass transition from nonlinear response experiments}}
	\par
	\null
	\par
	Caroline Crauste-Thibierge, Coralie Brun, Francois Ladieu, 
	
	Denis L'H\^ote, Giulio Biroli, Jean-Philippe Bouchaud

\end{center}
\vskip 10mm

For the sake of completness, we detail in the four sections hereafter: (i) the connexion between the third harmonics of the fundamental response and the nonlinear part of the response; (ii) what can be expected for the third harmonics response of a dielectric liquid where no (glassy) correlations are present; (iii) some additionnal information about our experimental setup; (iv) how the upper limit of the heating contribution was computed (Fig. 4 of the letter) and the reduction which can be expected from a more refined estimate.

\section{I) Nonlinear response measurements through third harmonics detection} 

As in the main paper, we consider the nonlinear response of a dielectric system to a time dependent electric field $E(t)$. The most general relationship between the polarisation $P(t)$ and the excitation $E(t)$ is a series expansion in $E$ \cite{SThibierge}. For a purely ac field, the even terms are forbidden because of the $E(t) \to -E(t)$ symmetry, which yields :
\begin{equation}
\frac {P(t)}{\epsilon_0} = \int_{-\infty}^{\infty}\chi_1(t-t')E(t')dt' + \int_{-\infty}^{\infty}\chi_3(t-t'_1,t-t'_2,t-t'_3) 
\times E(t'_1)E(t'_2)E(t'_3)dt'_1dt'_2dt'_3 + ...
\label{eqchidet}.
\end{equation}
In this equation $\epsilon_0$ is the dielectric constant of vacuum, $\chi_1$ the linear susceptibility and $\chi_3$ the cubic nonlinear susceptibility in the time domain. The dots in Eq.~\ref{eqchidet} indicate an infinite sum involving higher order nonlinear susceptibilities $\chi_5$, etc. Note that causality implies $\chi_{i}(t<0) =0$. The Fourier transform of Eq.~\ref{eqchidet} for a purely ac field $E(t)=E \cos(\omega t)$ gives:  
\begin{eqnarray}
\frac {P(\omega')}{\epsilon_0} &=& \frac{E}{2}\left[ \chi_1(\omega)+ \frac{3E^2}{4}\chi_3(-\omega,\omega,\omega)+... \right] \delta(\omega'-\omega) \nonumber \\
&+& \frac{E}{2}\left[ \chi_1(-\omega)+ \frac{3E^2}{4}\chi_3(\omega,-\omega,-\omega)+... \right] \delta(\omega'+\omega) \nonumber \\
&+& \frac{E^3}{8}\chi_3(\omega,\omega,\omega) \delta(\omega'-3\omega) \nonumber \\
&+& \frac{E^3}{8}\chi_3(-\omega,-\omega,-\omega) \delta(\omega'+3\omega)+... 
\label{eqchideomega} ,
\end{eqnarray}
where the polarization $P$ and the susceptibilities $\chi_i$ are now taken in the frequency domain and the dots indicate again infinite sums involving higher order terms.
The response $P(t)$ to $E(t)$ = $E \cos(\omega t)$ can thus be written 
\begin{equation}
P(t)/\epsilon_0 = Re \left[(E \chi_1(\omega)+3/4E^3 \chi_{\bar{3}}(\omega)+ ...) e^{-i\omega t}\right] 
+ Re \left[1/4E^3 \chi_3(\omega) e^{-i3\omega t} + ...\right]+ ....               
\label{eqchideomegapract1} 
\end{equation}
To obtain Eq.~(\ref{eqchideomegapract1}), we have used the fact that because $\chi_1$ and $\chi_3$ are real in the time domain, their Fourier transform verify 
$\chi_1^*(\omega)$ = $\chi_1(-\omega)$ and $\chi_3^*(\omega_1,\omega_2,\omega_3)$ = $\chi_3(-\omega_1,-\omega_2,-\omega_3)$ (the star denotes the complex conjugate), and the invariance of $\chi_3$ by permutation of its arguments. For simplicity, we write $\chi_3(\omega)$ = $\chi_3(\omega,\omega,\omega)$ and $\chi_{\bar{3}}(\omega)$ = $\chi_3(-\omega,\omega,\omega)$. Eq.~(\ref{eqchideomegapract1}) can be written 

\begin{multline}
P(t)/\epsilon_0 = E(\chi'_1 \cos\omega t + \chi''_1 \sin\omega t) + 3/4 E^3(\chi'_{\bar{3}} \cos\omega t + \chi''_{\bar{3}} \sin\omega t)+... \\
+1/4 E^3(\chi'_3 \cos3\omega t + \chi''_3 \sin3\omega t)+...,
\label{eqchideomegapract2} 
\end{multline}

where the susceptibilities $\chi_i$ are given as a function of their real and imaginary parts $\chi_i'$ and $\chi_i''$. For practical applications, the modulii and arguments $\left|\chi_i\right|$ and $\delta_i$ are rather used:

\begin{multline}               
P(t)/\epsilon_0 = E\left|\chi_1\right| \cos(\omega t - \delta_1) + 3/4 E^3\left|\chi_{\bar{3}}\right| \cos(\omega t - \delta_{\bar{3}}) + \\ 
+... + 1/4E^3\left|\chi_3\right| \cos(3\omega t - \delta_3)+...  
\label{eqchideomegapract3} 
\end{multline}

We see in the rhs in Eqs~\ref{eqchideomegapract1}-\ref{eqchideomegapract3} that the nonlinear susceptibility of interest, namely $\chi_3$ appearing in Eq. 1 of the letter, is \textit{directly given} by the measurement of the third harmonics of the polarisation. 

\section{II) Third harmonics for a liquid without glassy correlations.} 

We briefly summarize here what can be expected for the third harmonics in the case of a liquid without any glassy correlations. We use the work of D\'ejardin and Kalmykov \cite{SDejardin} which studies the nonlinear dielectric relaxation of an assembly of rigid polar symmetric particles. The particles are assumed to be diluted enough in a non polar solvent, so that any interaction between them can be safely neglected. A strong electric field is applied to the system, which biases the (noninertial) rotational Brownian motion of the polar particles and yields a net dielectric response. This response is expressed by expanding the relaxation functions as a Fourier series in the time domain. Then the infinite hierarchy of recurrence equations for the Fourier components is obtained in terms of a matrix continued fraction. 

We focus on the case where the field is purely ac, namely $E(t)=E \cos \omega t$. In the following equations, the subscript ``D" refers to the calculation of D\'ejardin and Kalmykov. By using their Eq. (21), one gets :
\begin{equation}
\frac{\chi_{3,D}(\omega)}{\chi_{3,D}(\omega=0)} = -3 \frac{3-17\omega^2 \tau_D^2+i \omega \tau_D(14-6 \omega^2 \tau_D^2)}{(1+\omega^2 \tau_D^2)(9+4\omega^2 \tau_D^2)(1+9\omega^2 \tau_D^2)} \ ,
\label{chi3normDejardin}
\end{equation}
where $\tau_D$ is the usual Debye time since one finds $\chi''_{1,D} \propto \frac{\omega \tau_D}{1+\omega^2 \tau_D^2}$, i.e. $\chi''_1$ is maximum for the frequency $f_D = 1/(2\pi \tau_D)$. By using Eqs. (7) and (19) of D\'ejardin and Kalmykov, one can calculate all the prefactors and obtain $\chi_{3,D}(\omega =0)$: 
\begin{equation}
\chi_{3,D}(\omega=0) = \frac{\epsilon_0 (\Delta \chi_1)^2}{k_BT N_0} \frac{1}{5}\ ,
\end{equation}
where $N_0$ is the volumic concentration of polar particles, and where, as in the letter, $\Delta \chi_1 = \chi'_{1,D}(0)-\chi'_{1,D}(\infty)$. To get a benchmark for the behavior of $\chi_{3, trivial}$ for the `no glassy correlations' case, we set $f_D=f_{\alpha}$ as well as $N_0=1/a^3$ with $a^3$ the molecular volume in glycerol. We thus obtain from the above equations:  
\begin{equation}
\chi_{3, trivial}(\omega) = \frac{\epsilon_0 (\Delta \chi_1)^2 a^3}{k_BT} \left(\frac{1}{5}\frac{\chi_{3,D}(\omega)}{\chi_{3,D}(\omega=0)} \right)\ .
\label{eqintermediaire}
\end{equation}

\begin{figure*}
\hskip -11mm
\includegraphics[scale=1.27,angle=0]{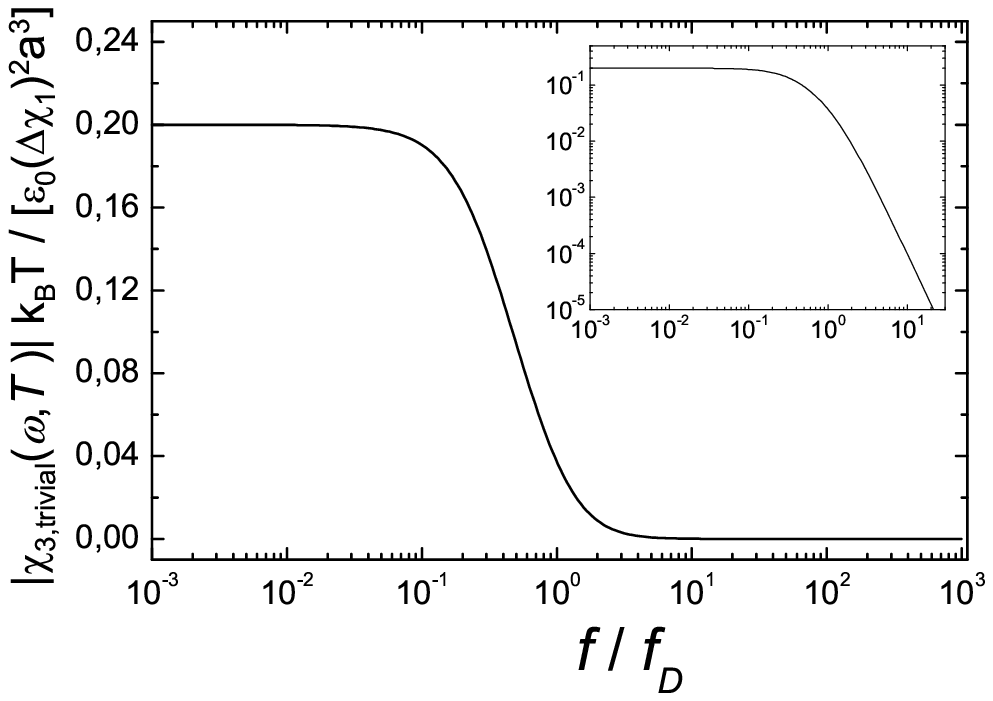}
\caption{\label{figS1} The quantity $\vert \mathnormal{X}_{trivial}\vert = \vert \frac{ \chi_{3, trivial}(\omega) k_BT}{\epsilon_0 (\Delta \chi_1)^2 a^3} \vert$ is plotted as a function of the ratio $f/f_{D}$. Inset: same graph in log-log plot showing that $\mathnormal{X}_{trivial}$ decreases as $f^{-3}$ when $f \gg f_{D}$.}
\end{figure*}

\begin{figure*}
\hskip -11mm
\includegraphics[scale=1.27,angle=0]{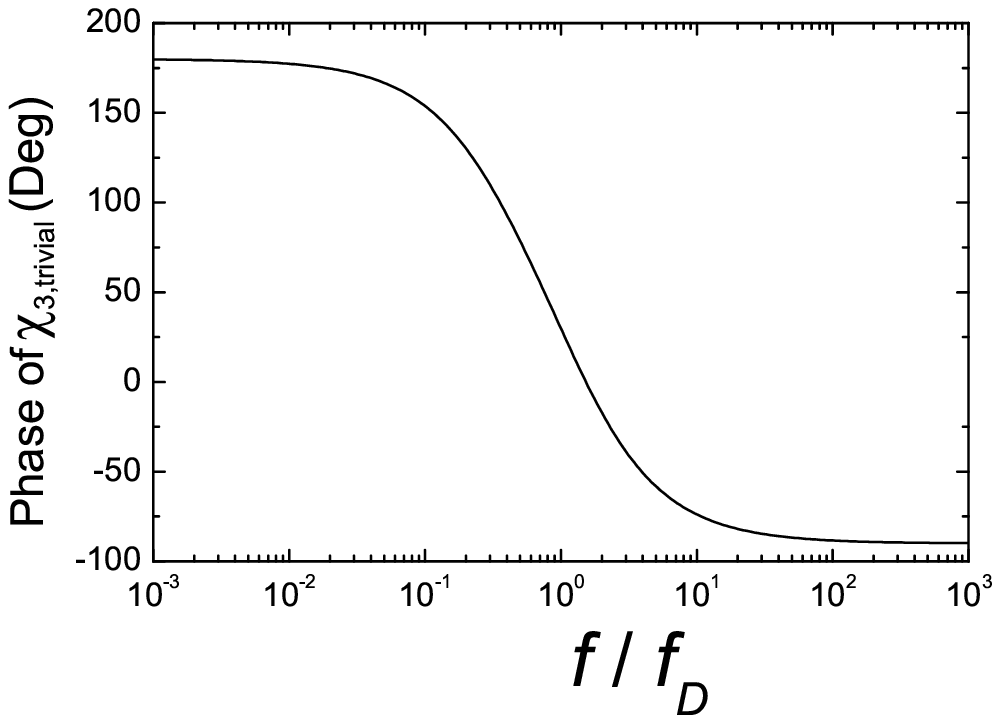}
\caption{\label{figS2} The phase of $\mathnormal{X}_{trivial}$ is plotted as a function of the ratio $f/f_{D}$.}
\end{figure*}

The last term in the rhs of the latter equation depends only on $f/f_D$, i.e. $f/f_{\alpha}$. Thus, the latter equation and Eq. (1) of the letter are formally similar: The quantity $\mathnormal{X}_{trivial} = \frac{ \chi_{3, trivial}(\omega) k_BT}{\epsilon_0 (\Delta \chi_1)^2 a^3}$ plays, for the `no glassy correlations' case, the same role as $\mathnormal{X}= \frac{ \chi_{3}(\omega) k_BT}{\epsilon_0 (\Delta \chi_1)^2 a^3}$ in the main letter. The behavior of $\mathnormal{X}_{trivial}$ is shown on Figs S1-S2. Three main features are noticeable:

(i) On Fig. S1, one sees that $\vert \mathnormal{X}_{trivial} \vert$ has no peak in frequency, at variance with what is observed on Figs (1),(2),(4) of the main letter. This strongly support that, as claimed in the letter, the peak in frequency of $\vert \mathnormal{X} \vert$ is directly linked to the effect of glassy correlations. 

(ii) From Eqs. (\ref{chi3normDejardin})-(\ref{eqintermediaire}), one sees that the magnitude of $\vert \mathnormal{X}_{trivial} \vert$ does not depend on the temperature $T$. The $T$ dependence of $\mathnormal{X}_{trivial}$ is only through $\tau_D(T)$, since we have set $f_{D}=f_{\alpha}$. This absence of temperature dependence of $\vert \mathnormal{X}_{trivial} \vert$ strongly supports our claim that the increase of the maximum value of $\vert \mathnormal{X} \vert$ seen on Figs (1),(3) of the letter comes from the variation of the correlation volume when the temperature is decreased towards $T_g$.

(iii) On Fig. S1, one sees that $\vert \mathnormal{X}_{trivial} \vert$ is maximum for $\omega =0$. The trivial part of $\chi_3$ thus dominates over the singular part in the limit of low frequencies $f \ll f_{\alpha}$. Of course the role of the trivial part of $\chi_3$ is larger when the non trivial part is lower, i.e. when the temperature is high. This naturally accounts for the departure to the scaling of $\mathnormal{X}$ as a function of $f/f_{\alpha}$ when both the temperature is high and the frequency $f$ is much lower than $f_{\alpha}$.

\section{III) More information about our experimental setup}

The principle of the experiment was that of a bridge including two
capacitors in which the dielectric layer was the supercooled liquid under study (here, glycerol) as described in section V of ref.~\cite{SThibierge}.
These two glycerol samples with different inter-electrode distances were placed in a closed cell filled with high purity glycerol (purity $> 99.6$ \%) in a dessicated atmosphere to prevent water absorption. The temperature of the cell was regulated by a LakeShore$^{\hbox{\textregistered}}$ temperature controller with a 50 Ohms heater fixed on the cell. The cell and its heater were placed in a cryostat in which the cold stage remained at $\sim$ 20 K and was connected to the cell through a thermal impedance. This device allowed a temperature regulation with a precision of $\sim 50$ mK, and a temperature quench from 300 to 200 K in $\sim$ 1.5 hour. 

The electrodes of the two capacitors are $6$ mm thick rods of high purity copper on which $200$ nm of gold was evaporated to prevent the formation of any oxydized layer. A highly specialised surface treatment reduced any surface defects: The final roughness was about 10 nm, while the planeity, i.e. the difference between the highest point of the electrodes and the mean plane, was better than $1\,\mu$m. 

Most of the results presented in the letter were obtained with Mylar$^{\hbox{\textregistered}}$ spacers depicted in the letter, giving inter-electrode distances of $19$ $\mu$m and $41$ $\mu$m for the two samples. For completeness, we have added a few results obtained very recently by using spacers made of resin patterned by photolithography (with a thickness $L$ $= 12$ $\mu$m for the thin sample and $25$ $\mu$m for the thick sample). 

The bridge was fed by a low distortion voltage source (Stanford Research Systems$^{\hbox{\textregistered}}$ DS360) which gave a rms excitation voltage $\leq$ 14V at a frequency $f$ = $\omega/2\pi$ ranging from 10 mHz to 200 kHz.  
The nonlinear susceptibility $\chi_{3}$ was determined after having carefully checked the cubic dependence of the measured signal at $3\omega$ with respect to the excitation at $1\omega$. Last, for a significant fraction of the frequencies and temperatures investigated here, we have checked the quantitative consistency between the $3 \omega$ signal measured in our ``two sample bridge'' and the one found with the ``twin T notch filter'' depicted in Ref.~\cite{SThibierge}, which allows to measure $\chi_3$ by a different method.

\section{IV) Heating effects calculations}
Let us consider a sample made of a supercooled liquid excited by an oscillating field $E \cos(\omega t)$. For small enough {\it E} values, the resulting polarisation $P_{lin}$ is linear and reads :
\begin{equation}\label{S1}
\frac{P_{lin}(t)}{\epsilon_0 E}= \chi_{1}' \cos(\omega t) +\chi_{1}'' \sin(\omega t).
\end{equation}
When $E$ is increased, a part of the nonlinear response comes from the dissipated electrical power, the volumic density of which is $p(t)=\frac{1}{2} \epsilon_0 \chi_{1}'' \omega E^2\left(1+\cos(2 \omega t - \phi) \right)$, with $\phi$ = $-\pi + 2\arctan(\chi_1''/\chi_1')$
(see e.g. Appendix of Ref.~\cite{SRichert08}). The resulting heat propagates towards a ``thermostat'' which in our case can be considered to be the copper electrodes (thickness 6 mm, diameter 20 mm). The resulting average sample temperature increase $\delta T(t)$ is
\begin{equation}\label{S2}
\delta T(t) = \delta T_{0} +\delta T_{2} \cos(2\omega t - \phi_2),
\end{equation}
where the mean dc temperature increase $\delta T_0$ is larger or equal than the ac one $\delta T_2$, thus at any time $\delta T(t) \ge 0$. 
$\phi_2$ is a phase shift related to heat transport. As our measurements give the nonlinear dielectric response averaged over the sample volume, $\delta T(t)$ in equation~(\ref{S2}) is the temperature increase averaged over the same volume. Using equations~(\ref{S1})-(\ref{S2}), we thus obtain for the nonlinear part of the susceptibility due to heating effects:
\begin{equation}\label{S3}
\frac{P(t) - P_{lin}(t)}{\epsilon_0 E}= \left( \frac{\partial \chi_{1}'}{\partial T} \delta T(t) \right) \cos(\omega t) +\left( \frac{\partial \chi_{1}''}{\partial T} \delta T(t) \right) \sin(\omega t).
\end{equation}
This is an upper limit on the contribution of heating to the nonlinear response, since we have assumed that $\delta T(t)$ induces instantaneously a modification of the susceptibility. As already advocated in Ref. \cite{SRichert08}, this is questionable, specially in what concerns the contribution of $\delta T_2(t)$ which should be averaged out because of the finite relaxation time $\tau$ of the system (see after eq. \ref{S7}). By using equations~(\ref{S2}-\ref{S3}), one gets :
\begin{eqnarray}
 \frac{P(t) - P_{lin}(t)}{\epsilon_0 E} & = & \left[\left(\delta T_0 + \frac{1}{2}\delta T_2 \cos(\phi_2)\right)\frac{\partial \chi_{1}'}{\partial T} + \frac{1}{2}\delta T_2 \sin(\phi_2)\frac{\partial \chi_{1}''}{\partial T}\right] \cos(\omega t)\nonumber \\
   &   & + \left[\left(\delta T_0 - \frac{1}{2}\delta T_2 \cos(\phi_2)\right)\frac{\partial \chi_{1}''}{\partial T}  + \frac{1}{2}\delta T_2 \sin(\phi_2)\frac{\partial \chi_{1}'}{\partial T}\right] \sin(\omega t)\nonumber \\
&   & + \left[ \frac{1}{2}\delta T_2 \cos(\phi_2)\frac{\partial \chi_{1}'}{\partial T} - \frac{1}{2}\delta T_2 \sin(\phi_2)\frac{\partial \chi_{1}''}{\partial T}\right] \cos(3\omega t)\nonumber \\
&   & + \left[ \frac{1}{2}\delta T_2 \sin(\phi_2)\frac{\partial \chi_{1}'}{\partial T} + \frac{1}{2}\delta T_2 \cos(\phi_2)\frac{\partial \chi_{1}''}{\partial T}\right] \sin(3\omega t).
\label{S4}
\end{eqnarray}
The four terms in the right hand side of equation~(\ref{S4}) give the nonlinear response of the system due to heating that we define in general by
\begin{equation}\label{S5}
 \frac{P(t) - P_{lin}(t)}{\epsilon_0 E} = (\delta \chi_{1}')_h \cos(\omega t) + (\delta \chi_{1}'')_h \sin(\omega t) + \frac{E^2}{4} \chi_{3,h}' \cos(3 \omega t)+ \frac{E^2}{4} \chi_{3,h}'' \sin(3 \omega t).
\end{equation}
In the right hand side of equation~(\ref{S5}) each of the four terms are proportional to $E^2$. $\chi_{3,h}'$  and $\chi_{3,h}''$ do not depend on $E$, while $(\delta \chi_{1}')_h$ and $(\delta \chi_{1}'')_h$ are proportional to $E^2$. Our notations have been chosen in order to remain as close as possible to the notations of refs~\cite{SRichert08, SRichert06, SRichert07a}. 

We now calculate the expression of $\delta T(t)$ that one has to introduce in equation~(\ref{S4}) in order to obtain the nonlinear response in equation~(\ref{S5}). We just consider the simple heating effect, when the dynamical heterogeneities are not considered:
The supercooled liquid is thus just characterized by its thermal conductivity $\kappa_{th}$ and its total specific heat $c$. As in ref.~\cite{SBirge86}, we consider that $c$ is frequency dependent due to the fact that the slow degrees of freedom cannot contribute to $c$ for frequencies much larger than $f_{\alpha}=1/\tau$. Let us define $(x,y)$ as the plane of our copper electrodes, with $z=0$ for the lower electrode and $z=L$ for the upper one. Due to their very high thermal conductivity and to their large thickness ($6$ mm), the two electrodes can be considered, to a very good approximation, as a thermostat. The temperature increase $\delta \theta (x,y,z,t)$ of the supercooled liquid at point $(x,y,z)$ and time $t$ thus vanishes for $z=0$ and $z=L$. As the diameter $D$ = 2 cm of the electrodes is typically one thousand times larger than $L$, we may consider that $\delta \theta$ does not depend on $(x,y)$. We obtain $\delta \theta(z,t)$ by solving the heat propagation equation:
\begin{equation}\label{S6}
c \frac{\partial \delta \theta(z,t)}{\partial t} = \kappa_{th} \frac{\partial^2 \delta \theta(z,t)}{\partial z^2} + p(t).
\end{equation}
By averaging spatially the solution of equation~(\ref{S6}), we obtain the $\delta T(t)$ to be used in equation~(\ref{S4}):
\begin{eqnarray}
 \delta T_0 & = & \frac{\epsilon_{0} \chi_{1}'' \omega E^2 L^2}{24 \kappa_{th}}  \nonumber \\
 \delta T_{2}(t) & = & \delta T_0 \frac{\cos(2 \omega t-\phi_2)}{\sqrt{1+(2 \omega \tau_{th})^2}},
\label{S7}
\end{eqnarray}
where $\tau_{th}= cL^2/(\kappa_{th} \pi^2)$ and $\phi_2 = \phi + \arctan(2 \omega \tau_{th})$. By using these expressions in equation~(\ref{S4}) and identifying it with equation~(\ref{S5}), we obtain the upper limit to the heating contribution to $\chi_3$ (dotted line in the Fig. 4 of the Letter).

We now move to the problem evoked above, namely the fact that the finite relaxation time $\tau$ of the dipoles 
which contribute to the dielectric susceptibility should average out the modification of this susceptibility due to the oscillation $\delta T_{2}(t)$, specially in the case $\omega \tau \ge 1$. As a consequence, the heating contribution to $\chi_3$ should be reduced by a factor $R(\omega \tau) \le 1$. For a precise calculation of $R(\omega \tau)$, one should replace Eq. (\ref{S3}) by a microscopic equation accounting for the dynamics of the dipoles in the case of a thermal bath where the temperature has an oscillating component, which is of great complexity. For a first estimate, we make two very simplifying assumptions: 

(i)  We assume that the dipoles have a Debye dynamics with a given characteristic time $\tau(T)$. This is a simplifying assumption in the sense that when $T$ is close to $T_g$, it is well known that $\chi_1(\omega)$ is ``stretched'' with respect to a simple Debye law. In fact the Debye dynamics holds only at much higher temperatures, where the molecular motions are independent of each other, which allows to describe the non inertial rotational Brownian motion by the Smoluchowski equation for the probability distribution function of the orientations of the dipoles in configuration space \cite{SDejardin,SLangevin}. After an ensemble averaging of this equation, one gets the well known Debye equation for the dynamics of the average polarisation \cite{SDejardin} :

\begin{equation} \label{S8}
\tau \frac{\partial P}{\partial t} + P = \epsilon_{0} \Delta \chi_1 E \cos(\omega t)
\end{equation}

(ii) we assume that the main effect of $\delta T_2(t)$ is to modulate in time the value of $\tau$ while leaving unchanged the (Debye) dynamics. This can be justified by the fact that the temperature oscillation modulates the viscosity $\eta$, thus the relaxation time $\tau$ which is proportional to $\eta$. Considering the heatings of eq. (\ref{S7}), $\tau(t)$ is now given by 

\begin{equation} \label{S9}
\tau(t) = \tau_{lin} + \left(\frac{\partial \tau_{lin}}{\partial T} \right) \delta T_0 + \left(\frac{\partial \tau_{lin}}{\partial T} \right) \delta T_2 (t)\; ,
\end{equation}
 
where $\tau_{lin}$ is the value of $\tau$ at zero field. In the following, $\delta \tau_2$ will denote the amplitude of the $2 \omega$ modulation of $\tau$ due to $\delta T_2(t)$ and corresponding to the last term of Eq. (\ref{S9}). Of course, using Eq. (\ref{S9}) for $\tau(t)$ assumes that $\delta T_2(t)$ instantaneously fully affects $\tau$. A thorough modelization of this problem could lead to a more involved expression where $\delta \tau_2$ should be weaker than in the above expression. As we shall find that the third harmonics is proportionnal to $\delta \tau_2 /\tau_{lin}$, see below Eq. (\ref{S11}), we are lead to the conclusion that our new estimate should, again, be slightly overstimated.

We now insert $\tau(t)$ in Eq. (\ref{S8}) and set :
\begin{equation} \label{S10}
P(t) = P_{lin} \cos(\omega t - \Psi_{lin}) + \delta P_1 \cos(\omega t - \Psi_1) + P_3 \cos(3\omega t - \Psi_3) + ...
\end{equation}
where $P_{lin}, \Psi_{lin}, \delta P_1, \Psi_1, P_3, \Psi_3$ are to be determined. As we are only interested in the onset of nonlinear effects, $P_{lin}\propto E$ is much larger than $\delta P_1 \propto E^3$ and than  $P_3 \propto E^3$. This allows to neglect higher order harmonics (denoted by the dots in Eq. (\ref{S10})) and to resolve Eq. (\ref{S8}) by identification of the terms which have the same frequency and the same power of $E$. This yields:
\begin{equation}\label{S11}
P_3 = \frac{\epsilon_{0} \Delta \chi_1 E}{2}\frac{\delta \tau_2}{\tau_{lin}} \frac{\omega \tau_{lin}}{\sqrt{1+(\omega \tau_{lin})^2}\sqrt{1+(3\omega \tau_{lin})^2}} 
\end{equation}
where $\delta \tau_2$ denotes the amplitude of the $2 \omega$ modulation of $\tau$ arising from $\delta T_2(t)$ in the last term of Eq. (\ref{S9}): thus $\delta \tau_2 \propto E^2$, see Eq. (\ref{S7}), which yields the expected $P_3 \propto E^3$. 

We now have to compare this result to that obtained if we start from Eq. (\ref{S3}) and use a Debye linear susceptibility. A straightforward calculation shows that in that case the modulus of the third harmonics of the polarisation is given by the expression of Eq. (\ref{S11}) divided by a function $R(\omega \tau)$ given by :
\begin{equation}\label{S12}
R(\omega \tau) =  \frac{\sqrt{1+(\omega \tau_{lin})^2}}{\sqrt{1+(3 \omega \tau_{lin})^2}}
\end{equation}
As expected $R(\omega = 0)=1$ and $R(\omega \tau \gg 1)$ is smaller than $1$. We note that $R(\omega \tau \to \infty) =1/3$, which comes from the fact that $\tau(t)$ enters in Eq. (\ref{S8}) as a factor of $\partial P/ \partial t$: this gives a weight $3 \omega \tau$ contrarily to the case where one starts from Eq. (\ref{S3}) where this weight is simply $\omega \tau$. This reduction of the effect of $\delta T_2(t)$ on the polarisation can be seen as a first estimate of the fact that the dipoles average out the temperature oscillations. Of course, one could build a much more thorough model of this effect, but the reduction given by Eq. (\ref{S12}) is enough to ensure that the power law regime reported in the main letter is not affected by the heating contribution calculated here. Finaly, the lengths of the arrows of the Fig. 4 of the main letter are those predicted by Eq. (\ref{S12}).


\begin{thebibliography}{99}
\bibitem{Dyre} J.C. Dyre, Rev. Mod. Phys. {\bf 78}, 953 (2006).
\bibitem{Debenedetti}{P.G. Debenedetti, F.H. Stilinger, Nature \textbf{410}, 259-267 (2001)}.
\bibitem{HoleBurning} M.D. Ediger, Annu. Rev. Phys. Chem. {\bf 51}, 99 (2000).
\bibitem{RevRichert} R. Richert, J. Phys.: Condens. Matter {\bf 14} R703 (2002).
\bibitem{RMN} U. Tracht {\it et al.}, Phys. Rev. Lett. {\bf 81}, 2727 (1998).
\bibitem{science} L. Berthier {\it et al.}, Science {\bf 310},  1797 (2005). 
\bibitem{JCP1} G. Biroli, {\it et al.}, J. Chem. Phys. {\bf 126}, 184503 (2007).
\bibitem{PRE} C. Dalle-Ferrier {\it et al.}, Phys. Rev. E {\bf 76}, 041510 (2007).
\bibitem{NoteJCP1} See the discussion on this prefactor (called $\chi_0$) following Eq. (18) in Ref. \cite{JCP1}. 
\bibitem{BY} K. Binder, A.P. Young, Rev. Mod. Phys. {\bf 58}, 801 (1986).
\bibitem{LLevy} L.P. L\'evy, Phys. Rev. B {\bf 38}, 4963 (1988).
\bibitem{BB-PRB} J.-P. Bouchaud, G. Biroli, Phys. Rev. B {\bf 72}, 064204 (2005). 
\bibitem{Tarzia} M. Tarzia, G. Biroli, A. Lef\`evre, J.-P. Bouchaud, J. Chem. Phys. {\bf 132}, 054501 (2010).
\bibitem{Supplement} for the definition of $\chi_3$, the `no glass correlation' case, the experiment,  the heating contribution of Fig. 4. : see EPAPS document, appended after this reference list (the EPAPS URL reference will be given when the paper is published).
\bibitem{perpignan} J. L. D\'ejardin, Yu. P. Kalmykov, Phys. Rev. E {\bf 61} 1211 (2000).
\bibitem{Thibierge} C. Thibierge, D. L'H{\^o}te, F. Ladieu, R. Tourbot, Rev. Scient. Instrum. {\bf 79}, 103905  (2008).
\bibitem{Lunkenheimer} P. Lunkenheimer, U. Schneider, R. Brand, A. Loidl, Contemp. Phys. {\bf 41}, 15 (2000).
\bibitem{Rossler} A. Kudlik {\it et al.}, J. Molec. Struct. {\bf 479}, 201 (1999).
\bibitem{Autres1w} P. K. Dixon {\it et al.}, Phys. Rev. Lett. {\bf 65}, 1108 (1990).
\bibitem{Wu} An attempt to measure $\chi_3$ of a glass former was reported in L. Wu, Phys. Rev B {\bf 43}, 9906 (1991), with negative  results, probably due to an experimental problem.
\bibitem{Richert07a} S. Weinstein, R. Richert, Phys. Rev. B {\bf 75}, 064302 (2007).
\bibitem{Bottcher} J. F. B\"ottcher, P. Bordewijk, {\it Theory of electric polarization} (Elsevier, Amsterdam, 1973), p. 289.
\bibitem{Lunkenheimer2} P. Lunkenheimer {\it et al.}, Phys. Rev. Lett. {\bf 77}, 318 (1996).
\bibitem{Richert08} W. Huang, R. Richert, Eur. Phys. J. B {\bf 66}, 217 (2008).
\bibitem{Richert06} R. Richert, S. Weinstein, Phys. Rev. Lett. {\bf 97}, 095703 (2006).
\end{thebibliography}

\begin{thebibliography}{99}
\bibitem{SThibierge} Thibierge, C., L'H{\^o}te, D., Ladieu, F.  and Tourbot, R., A method for measuring the nonlinear response in dielectric spectroscopy through third harmonics detection. {\it Rev. Scient. Instrum.} {\bf 79}, 103905  (2008).
\bibitem{SDejardin} D\'ejardin, J.L., Kalmykov, Yu.P. Nonlinear dielectric relaxation of polar molecules in a strong ac electric field: Steady state response. {\it Phys. Rev. E} {\bf 61}, 1211 (2000).
\bibitem{SRichert08} Huang, W. and Richert, R., On the harmonic analysis of non-linear dielectric effects. \textit{Eur. Phys. J. B} {\bf 66}, 217 (2008).
\bibitem{SRichert06} Richert, R. and Weinstein, S., Nonlinear Dielectric Response and Thermodynamic Heterogeneity in Liquids. \textit{Phys. Rev. Lett.} {\bf 97}, 095703 (2006).
\bibitem{SRichert07a} Weinstein, S. and Richert, R., Nonlinear features in the dielectric behaviour of propylene glycol. \textit{Phys. Rev. B} {\bf 75}, 064302 (2007).
\bibitem{SBirge86} Birge, N. O., Specific heat spectroscopy of glycerol and propylene-glycol near the glass-transition. \textit{Phys. Rev.B} {\bf 34}, 1631 (1986).
\bibitem{SLangevin} The Debye dynamics can be also obtained by starting with the non inertial Langevin equation for the roational Brownian motion of a particule, by appropriate transformation of the variables and direct averaging of the stochastic equation so obtained.

\end{thebibliography}
\end{document}